\documentstyle[preprint,aps]{revtex}
\begin{document}
\draft
\title{Specific heat of an $S$=1/2 Heisenberg ladder compound 
Cu$_{2}$(C$_{5}$H$_{12}$N$_{2}$)$_{2}$Cl$_{4}$ under magnetic fields}
\author{M. Hagiwara and H. A. Katori}
\address{RIKEN (The Institute of Physical and Chemical Research), Wako, 
Saitama 351-0198, Japan}
\author{U. Schollw\"{o}ck}
\address{Sektion Physik, Ludwig-Maximilians-Universit\"{a}t 
M\"{u}nchen, Theresienstr. 37, 80333 M\"{u}nchen, Germany}
\author{H. -J. Mikeska}
\address{Institut f\"{u}r Theoretische Physik, Universit\"{a}t 
Hannover, Appelstr. 2, D-30167 Hannover, Germany}
\date{Received \date }
\maketitle
\begin{abstract}
Specific heat measurements down to 0.5 K have been performed on a 
single crystal sample of a spin-ladder like compound 
Cu$_{2}$(C$_{5}$H$_{12}$N$_{2}$)$_{2}$Cl$_{4}$ under magnetic fields up 
to 12 T.  The temperature dependence of the observed data in a magnetic 
field below 6 T is well reproduced by numerical 
results calculated for the $S$=1/2 two-leg ladder with 
$J_{\rm{rung}}$/$J_{\rm{leg}}$=5.   In the gapless region above 7 T 
($H_{\rm{c1}}$), 
the agreement between experiment and calculation is good above about 2 
K and a sharp and a round peak were observed below 2 K in a magnetic 
field around 10 T, but the numerical data show 
only a round peak, the magnitude of which is smaller than that of the observed 
one.   The origin of the sharp peak and the difference between the 
experimental and numerical round peak are discussed. 
\end{abstract}
\section{Introduction}
Recently, there has been a considerable interest in quantum spin 
systems with a spin gap above the singlet ground state.   One of the 
examples studied extensively are one dimensional 
Heisenberg antiferromagnets (1DHAFs) with integer spin values, 
especially spin($S$) one, which is associated with Haldane's 
prediction\cite{haldane}.  Now, these studies extend those of the $S$=1 
antiferromagnetic bond 
alternating chains\cite{hagiwara1,narumi}.  Another case is 
$S$=1/2 two leg spin-ladder which is investigated as an intermediate system 
between one- and two-dimensional systems\cite{rice,azuma}.   Spin-ladder 
systems have been studied in relation to the Haldane problem on one hand\cite{hida,watanabe} and high T$_{c}$ superconductivity on the other 
hand\cite{dagotto,uehara}.   Most of the spin-ladder systems 
investigated so far are copper oxides\cite{azuma,matsuda}.   One of 
remarkable things is that a 
lightly hole doped two leg ladder exhibits superconductivity under 
high pressure\cite{uehara,nagata} as expected 
theoretically\cite{rice}.   On the other hand, spin-ladder like 
copper complexes Cu$_{2}$(C$_{5}$H$_{12}$N$_{2}$)$_{2}$Cl$_{2}$( 
abbreviated as 
CHpC)\cite{hammer,chaboussant1,hagiwara2,chaboussant2,chiba,ohta,calemczuk,hagiwara3} and 
KCuCl$_{3}$ and its family compounds\cite{tanaka,takatsu,oosawa} have 
been also studied extensively.   These compounds except for 
NH$_{4}$CuCl$_{3}$\cite{takatsu} 
have the singlet ground state and become gapless at a certain magnetic 
field ($H_{\rm{c1}}$).  The features of this gapless region at low 
temperatures have attracted much interest because the field induced long range 
ordering (LRO) in TlCuCl$_{3}$\cite{oosawa} has been interpreted as a Bose-Einstein 
condensation of magnons\cite{nikuni}.  This matter was originally 
argued by Affleck\cite{affleck} that the ground state above $H_{\rm{c1}}$ 
may be regarded as a Bose condensate of the low energy boson.   
Besides, the dimensionality is expected to appear in the power-law 
dependences of thermodynamic quantities on temperature, when 
approaching the quantum critical point by application of a magnetic 
field\cite{elstner}.   Furthermore, in the quantum critical 
region above the LRO temperature, we can investigate the feature of 
Tomonaga-Luttinger(TL) liquid for a quasi one-dimensional 
antiferromagnet\cite{mila,giamarchi}.

The compound CHpC has been studied extensively for several years by various 
experiments such as magnetic 
susceptibility\cite{hammer,chaboussant1,hagiwara2}, 
magnetization\cite{hammer,chaboussant1,hagiwara2}, 
NMR\cite{chaboussant2,chiba}, ESR\cite{hagiwara2,ohta}, 
specific heat\cite{hammer,calemczuk,hagiwara3} and neutron scattering 
measurements\cite{hammer}.   From these 
experiments, CHpC has the singlet ground state with an excitation gap 
of about 10 K and the gap collapses at about 7 T ($H_{\rm{c1}}$) and the 
saturation field is about 13 T ($H_{\rm{c2}}$).   This compound 
has an advantage of study on the gapless spin liquid phase under magnetic 
fields, because $H_{\rm{c1}}$ and $H_{\rm{c2}}$ are easily accessible fields 
with a conventional superconducting magnet.   Specific heat 
measurements in a magnetic field were done up to 9 T by Hammer 
$\it{et}$ $\it{al}$.\cite{hammer} and up to 8.25 T by Calemczuk 
$\it{et}$ $\it{al}$.\cite{calemczuk}.   In the present paper, we extend the specific 
heat measurements under magnetic fields up to 12 T beyond the symmetry 
field $H_{\rm{sym}}$(=($H_{\rm{c1}}$+$H_{\rm{c2}}$)/2) and compare the experimental 
results with those of numerical and analytical calculations.           

The format used in this paper is as follows: In Sec.~II, experimental and 
theoretical details are described.  Experimental results of 
magnetic susceptibility and specific heat under magnetic fields($H$) are reported 
in comparison with some numerical calculations in 
Sec.~III.   A sharp and a round peak observed at low temperatures in 
the gapless region are discussed in Sec.~IV.  

\section{Experimental and numerical details}
Powder samples of CHpC were synthesized according to the method reported in 
Ref.[26].  Equimolar amounts of 1,4-diazacycloheptane 
(C$_{5}$H$_{12}$N$_{2}$) and CuCl$_{2}\cdot$2H$_{2}$O were dissolved 
in warm methanol (60$^{\circ}$C) for 1 hour and left at room 
temperature for two days.   Single crystals of CHpC were obtained by 
the slow evaporation method from a methanol solution of powder 
samples of CHpC.   We obtained samples with 2$\times$2$\times$1 
mm$^{3}$ in typical size.    

CHpC crystallizes in the monoclinic system 
and belongs to the $P$2$_1$/$c$ space group\cite{chiari}.   The lattice 
constants and $\beta$ angle at room temperature are $a$=13.406(3)\AA, 
$b$=11.454(2)\AA\/, $c$=12.605(3)\AA and $\beta$=115.01(2)$^{\circ}$.   Cu 
dimeric units are linked to the neighboring ones via hydrogen bonds 
along the [101] direction.   The upper panel of Fig.~1 
shows a schematic view of the chain like structure along [101] of CHpC.  
Broken lines in the upper panel show the hydrogen bonds and the 
ellipsoids around the copper atoms represent the 3d hole orbitals.    
In the lower panel of Fig.~1, possible pathways of the exchange 
interaction between the neighboring Cu$^{2+}$ spins are depicted.   
The exchange interaction on the broken pathway must be much weaker 
than those on the other pathways because of the configuration of 3d 
copper and 3p chlorine orbitals, thus considering this system as the 
$S$=1/2 two leg spin-ladder. 

Magnetic susceptibilities ($M$/$H$) were measured with a SQUID 
magnetometer (Quantum Design's MPMS2) in RIKEN.   Specific heat 
measurements down to 0.5 K under magnetic fields up to 12 T were 
performed with Mag Lab$^{HC}$ micro calorimeter (Oxford Instruments) installed at 
the same place.    The relaxation method was employed so that we used 
only one single crystal with 4.3 mg in weight.   Numerical calculations  
 were done by the temperature-dependent density-matrix-renormalization-group 
(DMRG) method for the $S$=1/2 two-leg Heisenberg 
spin-ladder with a spin Hamiltonian written as
\begin{equation} 
{\cal H}=J_{rung}\sum_{j=1}^{N/2}{{\bf S}_{1,j}\cdot{\bf S}_{2,j}}+
J_{leg}\sum_{i=1}^{2}\sum_{j=1}^{N/2}{{\bf S}_{i,j}\cdot{\bf 
S}_{i,j+1}}-g\mu_{\rm{B}}H\sum_{i,j}{S_{i,j}^{z}}, 
\end{equation}
where $J_{rung}$ and $J_{leg}$ are the exchange constants along the rung and the leg, 
respectively and ${\bf S}_{i,j}$ an $S$=1/2 spin on the $i$-th leg, 
the $j$-th rung, $g$ the $g$-value of copper, $\mu_{\rm{B}}$ the Bohr 
magneton and $H$ the external magnetic field.   As we use a transfer 
matrix approach, it is free of finite size effects.    

\section{Experimental results and comparison with calculations}

\subsection{Magnetic susceptibility}
Solid circles of Fig.~2 shows the temperature dependence of magnetic susceptibility 
of a single crystal sample of CHpC along the chain direction.   A 
hump which is typical of low dimensional antiferromagnets is observed 
around 10 K and the susceptibility steeply decreases with further 
decrease of temperature toward zero Kelvin.   No increase of the 
susceptibility due to magnetic impurity or crystal defect is observed 
in this sample.   The solid line in this figure is a fit of the 
calculated susceptibilities to the experimental ones with the fitting 
parameters of $g$=2.1, $J_{\rm{rung}}$=13.1 K and $J_{\rm{leg}}$=2.62 K 
($J_{\rm{rung}}$/$J_{\rm{leg}}$=5).   The agreement between numerical and 
experimental susceptibilities is excellent.   Evaluated values and the 
ratio of the exchange constants are close to those estimated by other 
groups\cite{hammer,chaboussant1}.   In the following comparison of 
specific heat, we use the same fitting parameters.       

\subsection{Specific heat at $H$=0 T}
In Fig.~3, filled circles represent raw specific heat data.   Usually 
the lattice part of the specific heat is evaluated from the specific 
heat of a nonmagnetic isomorphous compound.   But no isomorphous 
compound with nonmagnetic atom like Zn exists.    In order 
to get the magnetic contribution of the specific heat, we subtract the 
lattice part ($C_{\rm{lattice}}$) of the specific heat from the raw data in the standard way,  
assuming  $C_{\rm{lattice}}$$\sim$$T^{3}$ at low temperatures.   The broken 
line in Fig.~3 is the lattice part of the specific heat with a 
coefficient of 0.004 (J/K Cu-mol) which is not far from those 
evaluated in other copper complexes ($\it{e.g.}$ 
0.0048 for (CH$_{3}$)$_{2}$CHNH$_{3}$CuCl$_{3}$\cite{manaka}) .    Open circles show the subtracted 
results, namely, magnetic part of the specific heat and we see a 
round peak due to the short range ordering around 5 K.    The solid line 
represents the specific heat at $H$=0 T calculated with the same 
fitting parameters as the susceptibility fitting.   The calculated 
specific heat satisfactorily agrees with the observed one.    At 
the temperature below the round peak, an exponential decay due to a spin 
gap is observed.   We fit the specific heat data at low temperatures 
to an expression of $C_{\rm{mag}}$$\sim$$T^{-3/2}$exp(-$\Delta$/$T$) to evaluate the 
energy gap.   This expression is deduced in the low temperature limit 
for the strong coupling case ($J_{\rm{rung}}$/$J_{\rm{leg}}$=5) approximating 
the dispersion $J_{\rm{rung}}$+$J_{\rm{leg}}$cos$\tilde{q}$ where $\tilde{q}$ 
is the component of wave vector transfer along the chain\cite{hammer}.   The 
fitting result is drawn as the solid line in Fig.4 with the fitting parameter 
$\Delta$=10.9 K, which is very close to the value of 
$J_{\rm{rung}}$-$J_{\rm{leg}}$(=10.48 K).               
 
\subsection{Specific heat at $H\neq$0 T}
Filled symbols in Figs. 5(a), 5(b) and 5(c) show the magnetic specific 
heat data  for magnetic fields below $H_{\rm{c1}}$, between $H_{\rm{c1}}$ 
and $H_{\rm{sym}}$ and 
above $H_{\rm{sym}}$, respectively.   In Figs.5(a), 5(b) and 5(c), a hump is observed 
above 5 K at each designated magnetic field.   Moreover, we observed two peaks 
below 2 K near $H_{\rm{sym}}$ in Figs.5(b) and 5(c).   Details of this low 
temperature part are depicted in the upper panel of Fig.6.   We 
obviously see two peaks, a sharp and a round peak at the 
magnetic field above 8.5 T.   In this figure, we plot the specific 
heat data under 10.5T for simplicity.     Figure 7 shows the 
magnetic field dependence of the specific heat.   Two peaks are 
observed below 0.82 K (only one peak at 0.61 K because of our instrumental 
limitation), whereas no peak is observed at 0.87 K.   We plot these two 
peaks at low temperatures in Figs.5-7 in the plane of $H$ $\it{vs.}$ $T$ of 
Fig.~8.   Plotted points are almost symmetric at $H_{\rm{sym}}$($\sim$10 
T) for both peaks.   

Next, we compare the experimental data with the numerical ones.    
Solid, broken, dotted lines in each figure of Fig.5 represent the 
numerical specific heat data for the designated magnetic fields.   In 
Fig.5(a), the agreement between experimental and numerical specific 
heat is excellent over the entire temperature range up to 9 K.   In Figs.5(b) and 5(c), the 
agreement between experiment and calculation is good above 2 
K, while the large deviation is seen below about 2 K.  In the lower 
panel of Fig.~6, calculated specific heat data are displayed for the 
magnetic fields corresponding to those in the upper panel.   The tendency of 
the calculated round peak to shift with change of fields is similar to 
the observed one, but the magnitude of the peak is much smaller than that 
of the observed peak.   We show in Fig.9 the magnitude of the round 
peak as a function of magnetic field.   The field at the 
calculated maximum peak slightly  shifts to the lower side compared to the 
experimental one.    The magnitude of the calculated peak is almost a 
half of the observed one.   

Entropy of this sample is calculated from the specific heat data shown 
in the upper panel of Fig.10 for the designated magnetic field.   
Correspondingly, we show the calculated entropy for the corresponding 
magnetic field in the lower panel of Fig.10.   Similar tendency of the 
entropy is observed in both figures.   With increasing the 
magnetic fields,  the entropy becomes higher at low temperatures.   
This behavior is opposite to that in a paramagnet.          
    
\section{Discussion and conclusions}
First let us discuss the origin of the sharp peak observed in the 
experiment. Usually, the sharp peak is thought to be caused by magnetic
long range ordering (LRO)\cite{giamarchi}, 
but it is controversial for the case of a 
spin ladder; for example, it predicts a field dependence of the transition
temperature (camel-type structure) contrary to experimental observations 
(dromendary-type structure). Recently, Nagaosa and 
Murakami\cite{nagaosa} argued that a lattice instability in the spin 
ladder is expected to occur above 
$H_{\rm{c1}}$.  From their study, the lattice distortion occurs in the spin 
ladder at an incommensurate wave vector corresponding to the 
magnetization. 

For a comparison of these scenarios with our experiment, we note that the one 
of the experimental consequences of Nagaosa's and Murakami's scenario 
is that a gap should appear below the transition temperature 
($T$$_{\rm{c}}$) and the specific heat should behave as 
$C\sim$exp(-$\gamma$$v_{0}$/$T$), 
where $\gamma$$v_{0}$ is the gap of the order 
of $T_{\rm{c}}$.    In the upper panel of Fig.11, we fit the experimental 
data at 10 T below the temperature at the sharp peak to this equation 
and obtained fairly good agreement between them with 
$\gamma$$v_{0}$=2.58 K, which is however quite far from the temperature 
$T_{\rm{c}}$ 
of about 0.8 K at 10 T. If the sharp peak is on the other hand
caused by antiferromagnetic long range order, the magnetic part of the 
specific 
heat should go as $T^{3}$ due to the antiferrmagnetic magnons. We 
show the fitting result using this equation in the lower part of 
Fig.11.   The agreement between experiment and calculation is 
comparably good to the previous fitting, when we add a constant. This
negative contribution is however unexplained.   
Both interpretations are therefore not really
conclusive, and discrepancies remain
between the possible sources of the sharp peak and the specific
heat data, calling for the proposal of a new transition scenario.

Theoretically we can discuss the round peak based on a fermion 
representation. The round 
peak can be attributed to the low energy excitation structure in 
a gapless spin liquid (TL liquid). As pointed out in the related 
context of mixed spin 1 and $\frac{1}{2}$ by Kolezhuk $\it{et. 
al.}$\cite{kolezhuk}, the round peak arises from a spinon band in the spin 
liquid system.  They reduce the Hilbert space of the problem, keeping 
only the most important states per elementary cell. In the 
strong coupling case of the spin ladder, this amounts to keeping 
for each rung only the singlet and the lowest-energy triplet component 
around $H_{\rm{c1}}$, mapping them to an effective 
spin 1/2 chain\cite{mila,giamarchi}.      
Then the effective Hamiltonian is given by
\begin{equation} 
{\cal H}=J_{leg}\sum_{i=1}^{N}{ (s_{i}^{x}\cdot s_{i+1}^{x}+s_{i}^{y}\cdot 
s_{i+1}^{y}+1/2s_{i}^{z}\cdot s_{i+1}^{z})}
-(g\mu_{\rm{B}}H-J_{rung}-1/2J_{leg})\sum_{i=1}^{N}s_{i}^{z}, 
\end{equation}
where $s_{i}^{x}$, $s_{i}^{y}$ and $s_{i}^{z}$ are the $x$, $y$ and $z$ 
component of the effective 1/2 spin. 
When, for a qualitative consideration, the $s_i^z \cdot s_{i+1}^z$
term is neglected, the remaining free fermion hamiltonian can be
easily treated.
Then a splitting of the peak is expected to 
occur\cite{kolezhuk} 
due to contributions from two different spinon bands (particle
and hole band) for magnetic fields off the symmetry field
$H_{sym}$. In experiment and in numerical studies on the
full original Hamiltonian of the ladder, we however observe only
one peak, which is probably due to the large deviation of the
effective Hamiltonian from the exactly solvable xy-limit, 
making the analytical prediction of a peak split less stringent.   
       
Finally, the difference of the magnitude between the experimental and the 
numerical round peak has to be discussed. When calculating the specific 
heat, we neglected the diagonal exchange interaction denoted by $j$ in 
Fig.~1.  This interaction can possibly affect the nature of the 
specific heat at low temperatures, but it should not allow for an effect
of more than 50 percent as observed. There must be an additional source of 
entropy and disorder, probably related to the physics driving the three
dimensional transition, which at present can only be speculated about.
 
In conclusion, we have made specific heat meaurements on a single 
crystal sample of the $S$=1/2 Heisenberg spin-ladder 
compound Cu$_{2}$(C$_{5}$H$_{12}$N$_{2}$)$_{2}$Cl$_{2}$ under 
magnetic fields.  Experimental data are compared to the numerical 
ones calculated with a temperature dependent DMRG method for the 
$S$=1/2 spin-ladder and are well reproduced when $H<H_{\rm{c1}}$, but when 
$H>H_{\rm{c1}}$ and for temperatures below about 2 K the 
magnitude of the observed round peak differs by a factor of about 2. 
The numerical calculations give quasiexact
results for the 1D Hamiltonian considered in the whole temperature and
magnetic field range. The remaining discrepancies have therefore to be
due to effects beyond that Hamiltonian, probably
due to low-temperature 3D coupling and/or further degrees of
freedom (such as phonons), as also indicated by the 3D transition.
  
We have observed two peaks, a sharp and a round peak at low 
temperatures below 2 K around $H_{\rm{sym}}$. The origin of the former peak 
is still controversial and two possibilities are discussed: 
the antiferromagnetic LRO\cite{giamarchi} 
and the lattice instability above $H_{\rm{c1}}$ 
proposed by Nagaosa and Murakami\cite{nagaosa}. The latter peak probably comes 
from a spinon band in the effective Tomonaga-Luttinger liquid. A simple 
analytical fermion representation predicts moreover a weak splitting of the
round peak into two round peaks for magnetic 
fields off the symmetry field $H_{\rm{sym}}$, but experimental and 
numerical results show consistently only one peak. This shows that on the
analytical level a more elaborate analysis of the Hamiltonian is 
needed to describe the round peak in the 1D picture, while qualitatively
the emergence of a new low-temperature structure in the specific heat
is well captured.   
 
\acknowledgments  
We would like to thank K.\ Katsumata for access to the SQUID 
magnetometer and Mag Lab$^{HC}$ calorimeter system.   M. H. expresses 
his thanks to J.\ -P.\ Boucher, N.\ Nagaosa and 
K.\ Totsuka for fruitful discussions.  
This work was partially supported 
by a Grant-in-Aid for Science Research from the 
Japanese Ministry of Education, Science, Sports and Culture. Thanks 
are also due to the Chemical Analysis and the Molecular 
Characterization Units in RIKEN.

 \begin{figure}
 \caption{A schematic view of the chain structure of 
 Cu$_{2}$(C$_{5}$H$_{12}$N$_{2}$)$_{2}$Cl$_{4}$ and the exchange 
 pathways.   Broken lines represent hydrogen bonds and ellipsoides show 
 3d hole orbitals of copper.}  
 \label{fig1}
 \end{figure}
 
 \begin{figure}
 \caption{Temperature dependence of magnetic susceptibilities of a 
 single crystal of Cu$_{2}$(C$_{5}$H$_{12}$N$_{2}$)$_{2}$Cl$_{4}$ 
 along the chain direction ([101]).  The solid line represents calculated 
 susceptibilities for the $S$=1/2 Heisenberg spin-ladder with the 
 fitting parameters shown in the panel.}
 \label{fig2}
 \end{figure}
 
 \begin{figure}
 \caption{Specific heat as a function of temperature at $H$=0 T.   
 Filled circles are raw specific heat data and a broken line 
 represents the estimated lattice part of the specific heat.   Open 
 circles show the magnetic specific heat.   The solid line represents 
 the specific heat calculated for the $S$=1/2 spin ladder with the 
 same fitting parameters as the susceptibility.}
 \label{fig3}
 \end{figure}
 
 \begin{figure}
 \caption{Specific heat at $H$=0 T in the plane of ln[$C_{\rm{mag}}$$T^{3/2}$] vs. 
 1/$T$.   The solid line is the result of a fit to the equation 
 $\sim$$T^{3/2}$exp(-$\Delta$/$T$) with $\Delta$=10.9 K.}
 \label{fig4}
 \end{figure}
 
 \begin{figure}
 \caption{Temperature dependence of the specific heat under the 
 designated magnetic fields along [101] 
 direction.   The solid, broken and dotted lines are 
 calculated specific heat data for the corresponding magnetic 
 fields.}
 \label{fig5}
 \end{figure} 
 
 \begin{figure}
 \caption{Details of the specific heat as a function of temperature at low temperatures below 1.5 
 K above $H_{\rm{c1}}$.}
 \label{fig6}
 \end{figure} 
 
 \begin{figure}
 \caption{Magnetic field dependence of the specific heat along the 
 [101] direction for the designated temperatures.  Solid lines are 
 guides for the eye.}
 \label{fig7}
 \end{figure} 
 
 \begin{figure}
 \caption{Plot of the peaks of the specific heat at low temperatures 
 above 7 T in the magnetic field $\it{vs}$. temperature plane.   
 Filled circles and open squares correspond to the sharp peaks and the 
 round ones, respectively.  Solid lines are guides for the eye.}
 \label{fig8}
 \end{figure}
 
 \begin{figure}
 \caption{Magnitude of the round peak as a function of magnetic 
 field.  Filled circles and open squares represent the magnetic 
 specific heat data for experiment and calculation, respectively.  
 Solid lines are guides for the eye.}
 \label{fig9}
 \end{figure} 
 
 \begin{figure}
 \caption{(a) Entropy as a function of temperature calculated from the 
 observed specific heat data for designated magnetic fields.  The entropy at the lowest 
 temperature of about 0.5 K is evaluated as half of $C_{\rm{mag}}$ at the lowest 
 temperature. (b) Calculated entropy as a function of temperature 
 corresponding to the experiment.  Solid lines are guides for the 
 eye.}
 \label{fig10}
 \end{figure}
 
 \begin{figure}
 \caption{(a) Specific heat data at 10 T in the plane of 
 -ln$C_{\rm{mag}}$ 
 versus 1/$T$.   The solid line represents a fit of the experimental 
 data to an equation $\sim$exp(-$\gamma$$v_{0}$/T).   (b) Specific 
 heat data at 10 T in the plane of $C_{\rm{mag}}$ versus $T^{3}$.   The 
 solid line shows a fit of the experimental data to an equation 
 $a$+$bT^{3}$ where $a$ and $b$ are fitting parameters.}
 \label{fig11}
 \end{figure}     
 
 \end{document}